\documentstyle[12pt]{article}

\advance\topmargin by -5\baselineskip
\advance\textheight by 4\baselineskip
\begin{document}

\thispagestyle{empty} 
\begin{titlepage}
\begin{center}
{\Huge Wavefunction corrections and off-forward gluon distributions
in diffractive $J/\psi$ electroproduction}\\
\vspace{1.0in}
{\bf Pervez Hoodbhoy}   \\
Department of Physics    \\
Quaid-e-Azam University  \\
Islamabad, Pakistan. \\
   \end{center}
\begin{abstract} 
Diffractive production of $J/\Psi$ particles by virtual photons on a proton target
is studied with a view towards understanding two important corrections to the leading
order result. First, the effect of Fermi motion of the heavy quarks is studied by 
performing a systematic expansion in the relative velocity, and a simple correction 
factor is derived. This is considerably less than estimated previously. Second, since 
the kinematics necessarily requires that non-zero
momentum be transferred to the proton, off-forward gluon distributions are probed by
the scattering process. To estimate the importance of the off-forwardness, we 
compute, in leading order perturbation theory, the extent of deviation from the 
usual forward gluon distribution in a quark. 
\end{abstract}
\end{titlepage}
\newpage

\section{INTRODUCTION}

There is considerable interest in diffractive electroproduction of $J/\Psi $
mesons off protons at high energies because this process is important for
studying the gluon density in a proton at low values of Bjorken x. This
interest stems from the simplicity of the leading order QCD expression for
longitudinally polarized photons which was first derived by Ryskin \cite
{Ryskin1,Ryskin2}, \label{one} 
\begin{equation}
\frac{d\sigma }{dt}(\gamma ^{*}+P\rightarrow J/\Psi +P)=\frac{16\pi
^3M\Gamma _{ee}}{3Q^6}\alpha _s(\stackrel{\_\_}{Q^2})[\xi g(\xi ,\stackrel{%
\_\_}{Q^2})]^2
\end{equation}
where, $\stackrel{\_\_}{Q^2}=\frac 14(Q^2+M^2)$, $M$ is the $J/\Psi $ mass,
and $\Gamma _{ee}$ is the decay width into leptons. The above equation was
derived under the assumptions of $s\gg Q^2\gg M^2\gg t$, and that Fermi
motion of the quarks in the meson can be entirely neglected. It was further
supposed that the gluon density appearing in eq.\ref{one} is that which
would be measured in some inclusive hard process, i.e. that it corresponds
to the matrix element of gluon operators between states of equal momentum.

In this paper we shall examine the effect of relaxing two assumptions which
go into eq.\ref{one}. The first is to take into account the correction
arising from the Fermi motion of the $c\bar c$ pair. In the work of Brodsky
et al\cite{Brodsky} this motion is contained in the vector meson light-cone
wavefunction $\psi ^{\mbox{v}}(k_{\bot },x)$, a quantity which is in
principle calculable from lattice QCD but whose presently unknown form is an
important source of ambiguity. For example, Frankfurt et al\cite{Frankfurt}
conclude that wavefunction effects suppress $J/\Psi $ production by a factor
of 3 or more. However Ryskin et al\cite{Ryskin2} estimate a suppression
factor of $0.4\leq F^2\leq 0.6$. The detailed shape of the wavefunction
appears to be an important source of the difference.

The method of treating the diffractive process, as well as Fermi motion
corrections, used in this paper differs from previously used methods in an
essential way. Rather than work in the infinite momentum frame and in the $%
A^{+}=0$ gauge, we shall choose the rest frame of the $J/\Psi $ and the
Coulomb gauge for the soft gluons in the meson wavefunction. This is the
natural choice for heavy-quark systems because one can then use systematic
procedures, such as Non-Relativistic Quantum Chromodynamics\cite{Bodwin}
(NRQCD) or the method developed in refs. \cite{Hafsa1,Hafsa2,Yusuf}, in
order to evaluate quarkonium observables of interest to any desired level of
accuracy. However, for the gluons in the fast moving proton we shall
continue to use the $A^{+}=0$ gauge because this is the natural gauge to use
for parton distributions. The two kinds of gluons have very different
momenta and hence are effectively distinguishable, justifying the use of two
different gauges in two different parts of the same Feynman diagram. It
turns out that a gauge-invariant correction factor, derived in this paper,
multiplies eq.1, 
\[
\left( 1+\frac 89\!\frac{\nabla ^2\phi }{M^2\phi }\right) . 
\]
The second derivative of the wavefunction is understood to be evaluated at
the origin. It is a non-perturbative quantity whose value has to be inferred
from some other quarkonium processes, such as decays or production,
involving large momentum transfer. In previous work\cite{Hafsa1,Hafsa2} its
value was estimated, 
\[
\!\frac{\nabla ^2\phi }{M^2\phi }\approx -0.07 
\]
The correction factor due to Fermi motion is therefore around $0.96$, a
value considerably below the other estimates\cite{Ryskin2,Frankfurt}. Hence,
the ambiguity in extracting the normalization of the gluon distribution may
be under better control than anticipated so far.

The second issue to be considered in this paper is the gluon distribution
which appears in eq.1. Recently Ji\cite{Ji} identified certain twist-two
``off-forward'' quark distributions inside the proton which, when measured,
will reveal the orbital angular momentum content of the proton. Subsequently
Radyushkin\cite{Radyushkin} extended the discussion to the off-forward or
``asymmetric'' gluon distribution in the proton and pointed out that
diffractive vector meson electroproduction necessarily measures this
quantity. Here we have examined this issue further by considering gluon
radiation from a quark and explicitly computed the off-forward gluon
distribution in a quark to leading order in $\alpha _s$. This enables an
estimate to be made of the extent to which the gluon distribution measured
in $J/\Psi $ diffractive production differs from that which would be
measured in some inclusive process like $\gamma +P\rightarrow J/\Psi +X.$ 

\section{FERMI\ MOTION}

\subsection{Kinematics}

We consider a massless proton target and the $t=0$ limit. Define two null
vectors $p^\mu $ and $n^\mu $ with $p^2=n^2=0$ and $p\cdot n=1$, 
\begin{equation}
p^\mu =\frac \Lambda {\sqrt{2}}(1,0,0,1),\;\;n^\mu =\frac 1{\sqrt{2}\Lambda
}(1,0,0,-1)
\end{equation}
$p^\mu $ is also the proton's momentum. Although we shall not need to do so
explicitly, $\Lambda $ can be adjusted to bring the produced $J/\Psi $ to
rest. The kinematic region of interest is considered to be $s\gg Q^2\gg M^2$%
. With the definitions, 
\begin{equation}
\xi =-\frac{q^2}{2p\cdot q},\;\;q^2=-Q^2,
\end{equation}
the other momenta in Fig.1 are, 
\begin{eqnarray}
q^\mu &=&-\xi p^\mu +\frac{Q^2}{2\xi }n^\mu  \nonumber \\
P^{^{\prime \mu }} &=&p^\mu +\Delta ^\mu  \nonumber \\
\Delta ^\mu &=&-\xi p^\mu  \nonumber \\
K^\mu &=&\frac{\xi M^2}{Q^2}p^\mu +\frac{Q^2}{2\xi }n^\mu
\end{eqnarray}
The polarization vectors of the longitudinally polarized photon and $J/\Psi $
are, respectively, 
\begin{eqnarray}
\varepsilon _L^\mu &=&\frac \xi Qp^\mu +\frac Q{2\xi }n^\mu  \nonumber \\
E_L &=&-\xi \frac M{Q^2}p^\mu +\frac{Q^2}{2M\xi }n^\mu
\end{eqnarray}
These obey $\varepsilon_{{L}} . \varepsilon _{{L}}=1$, $\varepsilon _{{L}} .
q_{{L}}=0$, and $E_{{L}} . E_{{L}} = -1$, $E_{{L}} . K_{{L}} = 0$ with $%
K^2=M^2$. We have kept only leading terms and set $\Delta _{\bot }\approx 0$.

\subsection{Diagrams}

The leading order contribution to $J/\Psi $ diffractive production is given
by the sum of the diagrams shown in Fig.2, to which must be added the
contribution of two other diagrams that give the same numerical values
because of time-reversal symmetry. Consider, by way of example, the first of
these which has the expression, 
\begin{eqnarray}
{\sl A}_1 &=&\int \frac{d^4k}{(2\pi )^4}\frac{d^4\ell }{(2\pi )^4}Tr[S_{\mu
\nu }(k,\Delta )\,H_1^{\mu \nu }(k,\ell )M(\ell )],  \nonumber  \label{amp}
\\
H_1^{\mu \nu }(k,\ell ) &=&e_Q\ g^2\;\gamma ^\mu S_F(k+q-K/2+\ell )\gamma
^\nu S_F(q-K/2+\ell ){\not \!\epsilon (q)}.
\end{eqnarray}
The perturbative part $H^{\mu \nu }(k,\ell )$ is different for the other
diagram but the other factors in eq.\ref{amp} remain unchanged. We have not
indicated colour explicitly in the above; its inclusion will amount to a
simple factor which will be inserted at the end of the calculation. The
non-perturbative information of the vector meson is contained in the
Bethe-Salpeter wavefunction $M(\ell )$, 
\begin{equation}
M(\ell )=\int d^4x\;e^{i\ell \cdot x}\langle K,E|T[\psi (x/2)\bar \psi
(-x/2)]|0\rangle .  \label{BS}
\end{equation}
In the above, $\ell ^\mu $ and $x^\mu $ are, respectively, the relative
momentum and relative distance of the $c\bar c$ pair. The non-perturbative
information of the gluons in the proton is contained in $S^{\mu \nu }$, 
\begin{equation}
S^{\mu \nu }(k,\Delta )=\int d^4x\;e^{i(k+\frac 12\Delta )\cdot x}\langle
P^{^{\prime }}|T[A^\mu (-x/2)A^\nu (x/2)]|P\rangle
\end{equation}
While the diagrams in Fig.2 contain the leading order contribution to the
cross-section, they also contain parts which are next to leading order
(NLO). The sense in which these are to be understood as ``higher order''
will be made precise later. Other diagrams will have to be included (see
Fig.3) for a complete calculation at the NLO level.

\subsection{Expansion}

The diffractive process considered here has two large scales, $Q^2\gg M^2\gg
\Lambda _{QCD}^2$. Since a $c\bar c$ system is close to being a
non-relativistic coulombic bound-state, it allows for an expansion in powers
of the heavy quark relative velocity. Hence it is useful to expand the inner
integral in eq.\ref{amp}, 
\begin{eqnarray}
\Omega (k) &=&\int \frac{d^4\ell }{(2\pi )^4}H^{\mu \nu }(k,\ell )M(\ell
)=\sum_{n=0}^\infty \Omega _n^{\mu \nu }  \nonumber \\
\ &=&\sum_{n=0}^\infty \frac 1{n!}\frac \partial {\partial \ell ^{\alpha
_1}}\cdot \cdot \cdot \frac \partial {\partial \ell ^{\alpha _n}}H^{\mu \nu
}\!\mid _{\ell =0}M^{\alpha _1\cdot \cdot \cdot \alpha _n}  \label{exp}
\end{eqnarray}
where, 
\begin{eqnarray}
M^{\alpha _1\cdot \cdot \cdot \alpha _n} &=&\int \frac{d^4\ell }{(2\pi )^4}%
\ell ^{\alpha _1}\cdot \cdot \cdot \ell ^{\alpha _n}M(\ell )  \nonumber
\label{me} \\
\ &=&i\partial ^{\alpha _1}\cdot \cdot \cdot i\partial ^{\alpha _n}\langle
K,E|T[\psi (x/2)\bar \psi (-x/2)]|0\rangle \mid _{x=0}  \label{M}
\end{eqnarray}

The set of constants $M^{\alpha _1\cdot \cdot \cdot \alpha _n}$ provide a
description equivalent to that of the original BS wavefunction in eq.\ref{BS}%
. The expansion eq.\ref{exp} is useful because the quarks are nearly on
mass-shell: $(\frac 12K+\ell )^2\approx m^2$ implies that all components of $%
\ell ^\mu $ are small relative to the quark mass $m$ in the meson's rest
frame and, in particular, $\ell \cdot n\sim (m/Q)^2$. In the large $Q^2$
limit this implies considerable simplification, giving a limit approximately
independent of $\ell $, 
\begin{eqnarray}
\frac 1{(k+q-K/2+\ell )^2-m^2+i\varepsilon } &\approx &\frac{2\xi }{Q^2}%
\frac 1{k\cdot n-\xi +i\varepsilon }, \\
\frac 1{(q-K/2+\ell )^2-m^2+i\varepsilon } &\approx &-\frac 2{Q^2}
\end{eqnarray}
Inclusion of Fermi motion requires that we keep a sufficient number of
derivatives w.r.t $\ell $ in eq.\ref{exp}. These may be computed using the
simple Ward identity, 
\begin{equation}
\frac{\partial}{\partial {\ell}_{\alpha}} S_F=-S_F {\gamma}^{\alpha} S_{%
{\small {F}}},
\end{equation}
and the $Q^2\rightarrow \infty $ limit should be taken after performing the
trace algebra. Stated in words, a differentiation of either propagator in eq.%
\ref{amp} w.r.t $\ell $ splits that propagator into two. Since we shall work
upto $O($v$^2)$, only two derivatives of $H^{\mu \nu }(k,\ell )$ are needed.

\subsection{Gauge Invariance}

It is obvious from the occurence of the ordinary derivatives in eq.\ref{M},
or the form of the BS wavefunction eq.\ref{BS}, that gauge invariance has
been violated. In earlier work on quarkonium processes\cite
{Hafsa1,Hafsa2,Yusuf} we have encountered an identical situation -- the
diagrams in Fig.2 yield expressions which are not gauge invariant to $O($v$%
^2)$ and one needs to consider additional diagrams, which are higher order
in $\alpha _s$. These are shown in Fig.3. The gluon fields indicated in
these diagrams combine with the ordinary derivatives to yield covariant
derivatives, $\partial ^\alpha \rightarrow D^\alpha ,$ thereby restoring
gauge invariance. In the Coulomb gauge, the contribution of explicit gluons
is $O($v$^3)$ and so the reduction of the Bethe-Salpeter equation performed
in ref.\cite{Keung} without explicit gluons is adequate upto $O($v$^2)$. We
therefore arrive at the following gauge-invariant matrix elements: 
\begin{eqnarray}
\langle K,E|\psi \bar \psi |0\rangle &=&\frac 12M^{1/2}\!\left( \phi \!+\!%
\frac{\nabla ^2\phi }{M^2}\right)\; \!\!\!\not \!\!\!\:E^{*}\left( 1\!+\!%
\frac{\not \!\!\!\:K}M\right)  \nonumber \\
&&\!-\frac 16M^{1/2}\!\frac{\nabla ^2\phi }{M^2}\;{\!}\!\!\not
\!\!\!\:E^{*}\left( 1\!-\!\frac{\not \!\!\!\:K}M\right)  \nonumber \\
\langle K,E|\psi \!\!\stackrel{\leftrightarrow }{iD_\alpha }\bar \psi
|0\rangle &=&\frac 13M^{3/2}\frac{\nabla ^2\phi }{M^2}E^{*\beta }(g_{\alpha
\beta }+i\epsilon _{\alpha \beta \mu \nu }\gamma ^\mu \gamma _5\frac{K^\nu }%
M)  \nonumber \\
\langle K,E|\psi {}\stackrel{\leftrightarrow }{iD_\alpha }\stackrel{%
\leftrightarrow }{iD_\beta }\bar \psi |0\rangle &=&\frac 16M^{5/2}\frac{%
\nabla ^2\phi }{M^2}\left( g_{\alpha \beta }-\frac{K_\alpha K_\beta }{M^2}%
\right) {{\!}\!\!\not \!\!\!\:E^{*}\left( 1\!+\!\frac{\not \!\!\!\:K}%
M\right) }.  \label{me}
\end{eqnarray}
In the above, $\phi $ and $\nabla ^2\phi $ are the non-relativistic
wavefunction and its second derivative evaluated at zero separation.
Inclusion of $\nabla ^2\phi $ amounts to taking the first step towards
inclusion of Fermi motion.

\subsection{Traces}

All the ingredients are now in place for calculating the trace of the quark
loops. Because we shall need only the leading twist piece, symmetric in $\mu 
$ and $\nu $, it will be sufficient to calculate, 
\begin{equation}
\Omega_n= Tr \sum_{{\small {i=1,2}}} \Omega_n^{{\small {ii}}}=(-g_{\mu \nu
}+p_\mu n_\nu +p_\nu n_\mu )Tr[\Omega _n^{\mu \nu }]
\end{equation}
for $n=0,1,2$ (n is the order of differentiation w.r.t $\ell $) and then
keep only the leading order term in $O(1/Q)$. We record below the results of
the calculation listing, for clarity, the relative contribution of only
those diagrams which give a non-zero contribution, 
\begin{eqnarray}
\Omega _0 &=&-\frac{4e_Q\ g^2\phi (0)}{M^{1/2}Q}\;2(1+2)\left( 1\!+\frac 23\!%
\frac{\nabla ^2\phi }{M^2\phi }\right) +O(1/Q^3)  \nonumber \\
\Omega _1 &=&O(1/Q^3)  \nonumber \\
\Omega _2 &=&\frac{4e_Q\ g^2\phi (0)}{M^{1/2}Q}\;2(\frac 23+\frac 43+\frac
43-\frac{8}3)\frac{\nabla ^2\phi }{M^2\phi }+O(1/Q^3)  \label{omeg}
\end{eqnarray}
The factor of 2 multiplying the brackets in the above equations comes from
the diagrams which are permutations of the ones shown. The sum over all
diagrams is, 
\begin{equation}
\Omega =-\frac{24e_Q\ g^2}{M^{1/2}Q}\!\phi (0)\left( 1+\frac 49\!\frac{%
\nabla ^2\phi }{M^2\phi }\right) +O(1/Q^3).  \label{omega}
\end{equation}
Note that this leading order contribution is in fact independent of the
gluon momentum $k$ in the $Q^2\rightarrow \infty $ limit. The term in the
brackets represents the correction due to the Fermi-motion of the heavy
quarks and its square is precisely the factor which modifies eq.1.

\section{GLUON\ DISTRIBUTION}

\subsection{Asymmetric Distribution}

Let us now return to the amplitude for diffractive scattering, a typical
contribution to which is given by eq.\ref{amp}. The photon and proton both
move along the $\hat z$ direction, and the gluons in the proton have limited 
$k_{\bot }^2$ and $k^2$. This means that one can perform a systematic
collinear expansion in these quantities just as in the treatment of
deep-inelastic scattering\cite{Ellis}, 
\begin{equation}
H^{\mu \nu }(k,\ell )=H^{\mu \nu }(k^{+},\ell )+(k-k^{+})_\alpha \partial
^\alpha H^{\mu \nu }(k,\ell )\mid _{k=k^{+}}+\cdot \cdot \cdot
\end{equation}
Keeping only the first, leading twist, term gives in the $A^{+}=0$ gauge, 
\begin{eqnarray}
&&\ \ \int \frac{d^4k}{(2\pi )^4}S_{\mu \nu }(k,\Delta )\,H^{\mu \nu
}(k,\ell )  \nonumber  \label{ccc} \\
\ &=&\int dy\int \frac{d\lambda \;}{2\pi }e^{i\lambda (y-\frac 12\xi
)}\langle P^{^{\prime }}|A_\mu (-\frac \lambda 2n)A_\nu (\frac \lambda
2n)]|P\rangle \;H^{\mu \nu }(y,\ell )  \label{ccc}
\end{eqnarray}
In the above we have set $x^{-}=\lambda n^{-}$ and $k^{+}=yp^{+}.$ The time
ordering operation becomes irrelevant on the light-cone.

The inner integral will now be analyzed following the discussion given by
Radyushkin\cite{Radyushkin}. Define the ``asymmetric distribution
function'', $F_\xi (X),$ as below, 
\begin{eqnarray}
&&\langle P^{^{\prime }}|n^{-}G^{+i}(-\frac \lambda 2n)n^{-}G^{+i}(\frac
\lambda 2n)]|P\rangle  \nonumber \\
\ &=&\frac 12\bar u(p^{\prime }){\!}{\!}n\,u(p)\int\limits_0^1dX\;\left\{
e^{i\lambda (X-\xi /2)}+e^{-i\lambda (X-\xi /2)}\right\} F_\xi (X)
\label{def}
\end{eqnarray}
A sum over transverse components ($i=1,2)$ is implied. The proton spinor
product is $\bar u(p^{\prime }){\!}{\!}n\,u(p)=2\sqrt{1-\xi }$, with the
initial and final protons having the same helicity and $p^{\prime }=(1-\xi
)p $. Making a Fourier transformation yields, 
\begin{equation}
\ \theta (y)F_\xi (y)+\theta (\xi -y)F_\xi (\xi -y)=\frac 1{\sqrt{1-\xi }%
}\int \frac{d\lambda \;}{2\pi }e^{i\lambda (y-\frac 12\xi )}\langle
P^{^{\prime }}|n^{-}G^{+i}(-\frac \lambda 2n)n^{-}G^{+i}(\frac \lambda
2n)]|P\rangle
\end{equation}
It is instructive to insert a complete set of states for $y>\xi >0$, 
\begin{equation}
\ F_\xi (y)\ =\frac{y(y-\xi )}{\sqrt{1-\xi }}\sum_k\delta (y-1+x)\langle
P^{^{\prime }}|A^i|k\rangle \langle k|A^i|P\rangle  \label{sum}
\end{equation}
Here $x=k\cdot n$ with $0<x<1$ is the momentum fraction carried by the
intermediate state. Comparing with the usual (diagonal) gluon distribution
function for $\xi =0$ it immediately follows that, 
\begin{eqnarray}
F_{\xi =0}(y) &=&y\,g(y),  \nonumber \\
g(y) &=&y\sum_k\delta (y-1+x)\langle P|A^i|k\rangle \langle k|A^i|P\rangle
\label{usual}
\end{eqnarray}

We shall now relate the matrix element in eq.\ref{ccc} to $F_\xi (y).$
Inverting the relation $G^{+i}=\partial ^{+}A^i$ gives, 
\begin{equation}
A^i(\lambda n)=n^{-}\int_0^\infty d\sigma \,G^{+i}(\lambda n+\sigma n).
\end{equation}
Inserting the above into eq.\ref{ccc} and using the definition of $F_\xi (y)$
in eq.\ref{def}, 
\begin{eqnarray}
&&\ \ \ \int \frac{d\lambda \;}{2\pi }e^{i\lambda (y-\frac 12\xi )}\langle
P^{^{\prime }}|\,A^i(-\frac \lambda 2n)A^i(\frac \lambda 2n)|P\rangle 
\nonumber \\
\ &=&-\sqrt{1-\xi }\left\{ \frac{F_\xi (y)}{y(\xi -y-i\varepsilon )}+\frac{%
F_\xi (\xi -y)}{(\xi -y-i\varepsilon )(y-i\varepsilon )}\right\}
\end{eqnarray}
The imaginary part of the above for $y>0$ is 
\[
-\pi \sqrt{1-\xi }\frac{F_\xi (\xi )}\xi \delta (\xi -y), 
\]
and hence, 
\begin{equation}
\ Im\int \frac{d^4k}{(2\pi )^4}\frac{d^4\ell }{(2\pi )^4}Tr[S_{\mu \nu
}(k,\Delta )\,H^{\mu \nu }(k,\ell )M(\ell )]=-\frac 12\pi \sqrt{1-\xi }\frac{%
F_\xi (\xi )}\xi \Omega \ .
\end{equation}

\subsection{Cross-section}

All the ingredients are now in place for calculating the cross-section for
the diffractive process under consideration, 
\begin{eqnarray}
\frac{d\sigma }{dt} &=&\frac 1{16\pi s^2}\left| A\right| ^2  \nonumber \\
\ &=&\frac 1{16\pi (Q^2/\xi )^2}\left( \frac 2{3\sqrt{3}}\right) ^2\left(
\frac 12\,\pi \sqrt{1-\xi }\frac{F_\xi (\xi )}\xi \Omega \right) ^2
\end{eqnarray}
The factor $\frac 2{3\sqrt{3}}$ comes from summing over colours, and $\Omega 
$ is the quantity calculated in the previous section, eq.\ref{omega} , from
the expansion of the heavy quark loop integral. Defining $\Gamma $ to be the
leading order decay width into lepton pairs, 
\begin{equation}
\Gamma =\frac{16\pi e_Q^2\alpha _e^2}{M^2},
\end{equation}
yields the following important result, 
\begin{equation}
\frac{d\sigma }{dt}=\frac{16\pi ^3M\Gamma }{3Q^6}\alpha _s(\stackrel{\_\_}{%
Q^2})\left[ \sqrt{1-\xi }F_\xi (\xi )\right] ^2\,\left( 1+\frac 89\!\frac{%
\nabla ^2\phi }{M^2\phi }\right) .
\end{equation}
Making the approximate identification, 
\begin{equation}
\sqrt{1-\xi }F_\xi (\xi )\approx \xi g(\xi ),  \label{zzz}
\end{equation}
and setting the last factor to unity reproduces eq.1 once again. This
identification was motivated by eq.\ref{usual} but the exact relation
between $F_\xi (\xi )$ and $g(\xi )$ is far from clear.

\subsection{Perturbative gluon distribution}

$F_\xi (\xi )$ and $g(\xi )$ can be known only if the non-perturbative
structure of the proton state is known. However, it would be highly
desirable to have at least some partial knowledge of their structure. To
this end, consider the following simple solvable problem: imagine that the
target proton is replaced by a single quark which can radiate a gluon. Its
light-cone wavefunction can be computed order by order in perturbation
theory, and the leading order matrix element is, 
\begin{equation}
\langle P^{^{\prime }}s^{^{\prime }}|\,A^i\,|ks\rangle =g\frac{k^{+}}{%
p^{^{\prime }+}}\frac 1{k_{\perp }^2}\sum_\lambda \bar u(p^{\prime
}s^{^{\prime }})\,\!{\!}\epsilon (l\lambda )\,u(ps)\,\varepsilon
{}^{*i}(l\lambda ),
\end{equation}
where $l,\lambda $ are the momenta and transverse polarizations of the
emitted gluon. Using, 
\begin{equation}
\sum_\lambda \varepsilon ^\mu (l\lambda )\,\varepsilon ^\nu {}^{*}(l\lambda
)=-g^{\mu \nu }+\frac{l^\mu n^\nu +l^\nu n^\mu }{l\cdot n},
\end{equation}
and summing over $i=1,2$ gives, 
\begin{equation}
\langle P^{^{\prime }}|A^i|k\rangle \langle k|A^i|P\rangle =g^2\frac{2x}{%
\sqrt{1-\xi }}\frac 1{k_{\perp }^2}\frac{1+x^2-\xi }{(1-x)(1-x-\xi )}.
\end{equation}
Inserting this into eq.\ref{sum} yields, 
\begin{equation}
\ \ F_\xi (y)=g^2\frac{2y(y-\xi )}{\left( 1-\xi \right) }\int \frac{d^2k}{%
16\pi ^3k_{\perp }^2}\ \int dx\delta (y-1+x)\frac{1+x^2-\xi }{(1-x)(1-x-\xi )%
}
\end{equation}
The last integral is both infrared and ultraviolet divergent. It is
regulated by inserting a low momentum scale cutoff $\mu \sim O(\Lambda
_{QCD})$ and a high momentum cutoff $k_{\perp }\sim O(Q).$ Multiplying by
the colour factor $C_F$ $=\frac 43$, we arrive at the perturbative {\it %
asymmetric }gluon distribution inside a quark, 
\begin{equation}
F_\xi (y)=\frac{2\alpha _s}{3\pi }\left\{ 1+\frac{(1-y)^2}{(1-\xi )}\right\}
\log \frac{Q^2}{\mu ^2}  \label{F}
\end{equation}
Note that, 
\begin{eqnarray}
\sqrt{1-\xi }F_\xi (\xi ) &=&\frac{2\alpha _s}{3\pi }\sqrt{1-\xi }\left\{
1+(1-\xi )\right\} \log \frac{Q^2}{\mu ^2}  \nonumber \\
\ &=&\frac{4\alpha _s}{3\pi }\left( 1-\xi +\frac 18\xi ^2+\cdot \cdot \cdot
\right)
\end{eqnarray}
but that the usual perturbative {\it symmetric} distribution, which can also
be obtained by first putting $\xi =0$ in eq.\ref{F} and then setting $y=\xi $
is, 
\begin{equation}
\xi g(\xi )=\frac{4\alpha _s}{3\pi }\left( 1-\xi +\frac 12\xi ^2\right) .
\end{equation}
Comparison of the last two formulae gives an estimate of the extent to which
the asymmetric distribution departs from the symmetric one as $\xi $ becomes
larger.

Finally, we remark that there exists some confusion in the literature about
various factors of two and four. First, it is claimed in the work of Brodsky
et al\cite{Brodsky} that the cross-section displayed in eq.1 must be
multiplied by $\frac 14$. We do not find this to be the case; the result of
Ryskin\cite{Ryskin1,Ryskin2} appears to be correct. This point has been
corroborated in ref.\cite{Frankfurt}. A second possible point of confusion
concerning the relation between $F_\xi (y)$ and the usual gluon distribution 
$g(\xi )$ has also now been resolved following the corrected definition
(which I have used in the final version of this paper) of $F_\xi (y)$ in ref.%
\cite{Radyushkin}. \newpage
\centerline{\bf Acknowledgements} I would like to thank Xiangdong Ji and
Daniel Wyler for a discussion. This work was supported in part by funds
provided by the Pakistan Science Foundation$.$

\begin{center}
{\Large Figure Captions}
\end{center}

\begin{tabbing}
\end{tabbing}
{\raggedright 1. Definition of kinematic variables for }$J/\Psi $
diffractive production off a proton target by a virtual photon. 
\begin{tabbing}
\end{tabbing}
{\raggedright 2. Diagrams which give non-zero contribution at order }$Q^{-1}$%
{\ and }v$^0.${\ The relative weight at this order a:b is as 1:2. Two other
diagrams, which are numerically equal by time-reveral invariance, are not
shown. The complete expression is given in eq.\ref{omeg}}. 
\begin{tabbing}
\end{tabbing}
{\raggedright 3. Diagrams which give non-zero contribution at order }$Q^{-1}$%
{\ and }v$^2.${\ The crosses denote connection to external gluons
originating from the proton. The relative weight at this order a:b:c:d:e:f
is as -1:1:-2:2:2:-4. Note that each internal gluon zero-momentum gluon
line, in the Coulomb gauge, is actually just a differentiation of the quark
propagator. Six other diagrams, which are numerically equal by time-reversal
invariance, are not shown. The complete expression is given in eq.\ref{omeg}}%
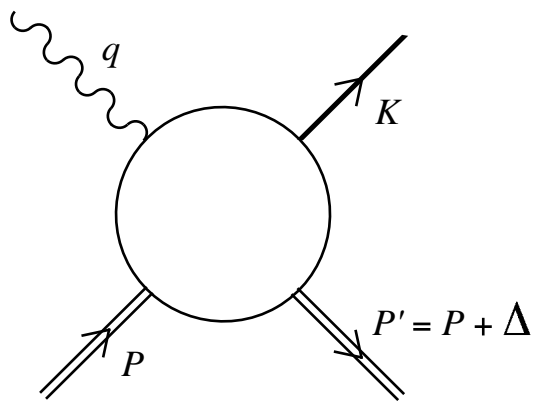
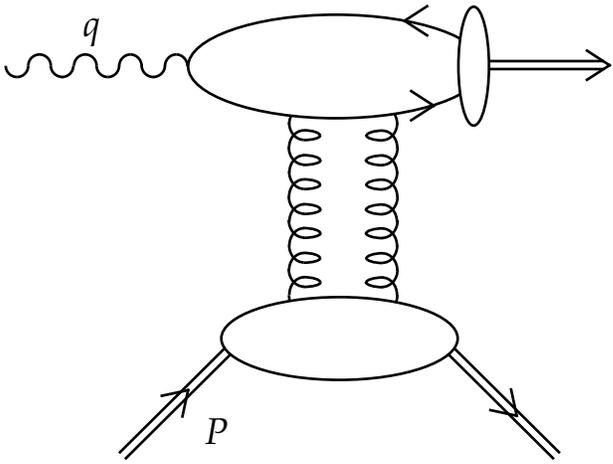
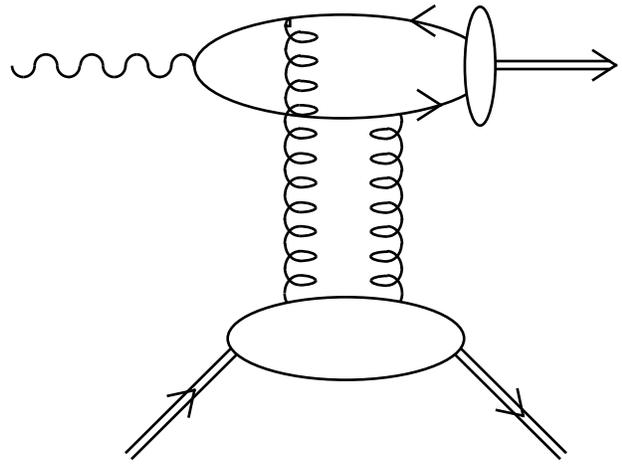
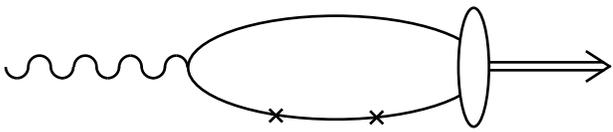
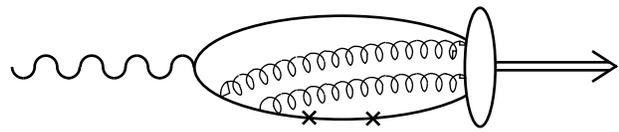
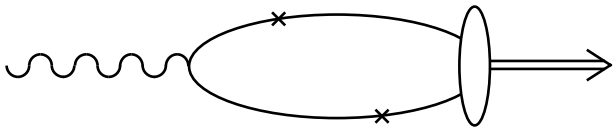
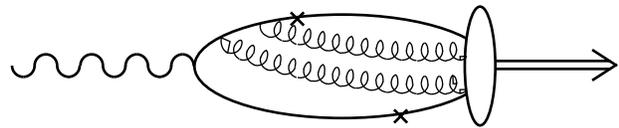
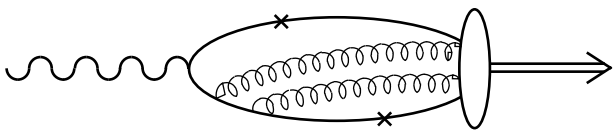
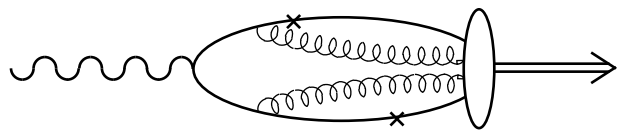
\end{document}